\documentclass[aip,jcp]{revtex4-1}

\usepackage{graphicx}

\draft 

\begin{document}

\title{Assessing computationally efficient isomerization dynamics:\\ $\Delta$SCF density-functional theory study of azobenzene molecular switching}

\author{Reinhard J. Maurer}
\author{Karsten Reuter}
\email[]{karsten.reuter@ch.tum.de}
\affiliation{Department Chemie, Technische Universit{\"a}t M{\"u}nchen, Lichtenbergstr. 4, D-85747 Garching, Germany}

\date{\today}

\begin{abstract}
We present a detailed comparison of the S0, S1 ($n \rightarrow \pi^*$) and S2 ($\pi \rightarrow \pi^*$) potential energy surfaces (PESs) of the prototypical molecular switch azobenzene as obtained by $\Delta$-self-consistent-field ($\Delta$SCF) Density-Functional Theory (DFT), time-dependent DFT (TD-DFT) and approximate Coupled Cluster Singles and Doubles (RI-CC2). All three methods unanimously agree in terms of the PES topologies, which are furthermore fully consistent with existing experimental data concerning the photo-isomerization mechanism. In particular, sum-method corrected $\Delta$SCF and TD-DFT yield very similar results for S1 and S2, when based on the same ground-state exchange-correlation (xc) functional. While these techniques yield the correct PES topology already on the level of semi-local xc functionals, reliable absolute excitation energies as compared to RI-CC2 or experiment require an xc treatment on the level of long-range corrected hybrids. Nevertheless, particularly the robustness of $\Delta$SCF with respect to state crossings as well as its numerical efficiency suggest this approach as a promising route to dynamical studies of larger azobenzene-containing systems.
\end{abstract}

\pacs{}

\maketitle 

\section{Introduction}
\label{intro}

A future molecular nanotechnology has to rely on techniques and systems that enable us to selectively interact with single molecules and prepare them in defined states. Only then molecules can act as well-defined building blocks for nanotechnological devices, e.g. for highly integrated data storage \cite{Liu1990,Gindre2006} or in photomechanical machines \cite{Ichimura2000,Hugel2002,Norikane2004}. A class of molecules that fulfills these requirements in solution are so-called molecular switches, i.e. molecules that can be reversibly switched between two or more stable states via external stimuli \cite{Ferringa2001}. In the quest to exploit this intriguing functionality for nanotechnological devices, research has recently concentrated on the properties of such molecules when they are localized at solid surfaces \cite{Browne2009a,Morgenstern2011}. Herein, adsorption at metal surfaces represents a particularly appealing sub-topic, with couplings to the underlying substrate that go beyond the mere weak physisorption limit. Notwithstanding the constant danger of ultrafast quenching of any (photo-)excitation due to a too strong coupling, the very intricacies of this interaction with the delocalized metal electron manifold is hoped to give rise to novel isomerization mechanisms and thereby to a switching behavior that is not attainable in gas-phase or solution. 

A corresponding switching has for instance been achieved for azobenzene derivatives at coinage metals \cite{Comstock2005,Comstock07,Comstock2008} and the geometric structure of the (meta)stable surface mounted molecular states is to some extent unraveled \cite{Tegeder2007,McNellis2009,McNellis2009a,Schmidt2010,Mercurio2010}. The observed non-trivial excitation wavelength, surface orientation \cite{Alemani2008,McNellis2009a} and functionalization \cite{Dokic2009,McNellis2010,McNellis2010a} dependence of the switching efficiency indeed suggests that the isomerization process proceeds either on molecular potential energy surfaces (PESs) that are strongly modified as compared to their established counterparts in solution, or even involves electronic excitations in the underlying 
metal \cite{Hagen2008,applphysa,Wolf2009}. In this situation independent insight as provided by material-specific first-principles theory would be highly desirable to further elucidate the mechanistic details of the switching function. Unfortunately, the very experimental evidence already indicates the tremendous challenge that metal-surface mounted switches pose to such modeling: On the one hand, the theory obviously needs to accurately describe both molecular ground and involved excited electronic states. Particularly the latter is commonly the realm of numerically highly-demanding correlated wave-function based approaches tractable only for very limited system sizes. On the other hand, the non-trivial influence of the substrate dictates its explicit treatment, which in order to properly describe the metal band structure needs to rely on extended supercell geometries. Together with the sheer lateral extension of flat-lying adsorbed molecules like azobenzene this gives rise to system sizes that are already at the cutting-edge of what can be tackled with approximate ground-state techniques like density-functional theory (DFT) with present-day semi-local exchange-correlation (xc) functionals \cite{McNellis2009,McNellis2009a,Mercurio2010,McNellis2010,McNellis2010a}. This calls for numerically highly efficient approaches to describe the excited states, which in fact should only impose CPU-costs comparable to those of a semi-local DFT ground-state calculation. 

It is self-evident that such approaches will be approximate in nature, and thus need to be carefully scrutinized to assess what can and what cannot be addressed reliably. A careful scrutiny requires accurate references as benchmark though. With only limited and indirect experimental information on excited PES topology available this primarily concerns higher-level theory. Notwithstanding, corresponding techniques might have their own limitations. As little conducive as uncritically applying approximate theories is then to readily dismiss them either for what they are or because of discrepancies with incorrect reference data. With this scope the objective of the present work is specifically to revisit the reliability of the numerically undemanding DFT-based $\Delta$ Self Consistent Field ($\Delta$SCF) approach to excited states \cite{Gunnarsson1976,Jones1989,Gorling1999} in the context of the prototypical molecular switch azobenzene (Ab, H$_5$C$_6$-N=N-C$_6$H$_5$). Apart from perfectly meeting the computational efficiency requirement this type of constrained DFT \cite{Wu2005,Behler2008} technique is particularly appealing as it is readily extended to applications at extended surfaces, even in case of appreciable adsorbate-surface hybridization \cite{Hellman2004,Gavnholt2008}. Unfortunately, a preceding study by Tiago {\em et al.} \cite{Tiago2005} on gas-phase azobenzene reported qualitatively different $\Delta$SCF PES topologies as compared to Time-Dependent DFT (TD-DFT) \cite{Casida2009} reference data, questioning the usefulness of this approach for an envisioned application to the isomerization mechanism of azobenzene (and its derivatives) at metal surfaces. 

Revisiting the problem, we employ approximate Coupled Cluster Singles and Doubles (RI-CC2) \cite{Christiansen1995,Hattig2000} as additional reference technique. Its accuracy in describing the lowest lying singlet excitations relevant for the isomerization of gas-phase azobenzene was already demonstrated by Fliegl {\em et al.} \cite{Fliegl2003}. For reasons of computational feasibility the present calculations are still exclusively performed for the free molecule in the gas-phase. Nevertheless, the discussion and assessment will also be geared towards the application of $\Delta$SCF to surface mounted azobenzene. As such, we e.g. concentrate on semi-local DFT with a gradient-corrected xc functional (GGA) \cite{Perdew1996} as ground-state basis for the approach, as such functionals are presently the unbeaten workhorse for metal surface studies. From the mapping of different ground- and excited-state PES cuts discussed in the context of azobenzene photoisomerization we arrive at the central conclusion that $\Delta$SCF, TD-DFT and RI-CC2 yield overall very similar PES topologies. The discrepancies reported before by Tiago {\em et al.} were caused by an undiscovered state crossing in their TD-DFT calculations. Limitations due to the approximate xc functional in $\Delta$SCF and TD-DFT primarily show up as global excited-state PES offsets, leading to a severe underestimation of absolute vertical excitation energies at e.g. the level of semi-local functionals.

\section{Computational Method}
\label{methods}

All ground-state calculations at the spin-polarized Kohn-Sham (KS) DFT level were performed with the all-electron full-potential DFT code FHI-aims \cite{Blum2009}. Centrally targeted are results as obtained with the GGA functional due to Perdew, Burke and Ernzerhof (PBE) \cite{Perdew1996} to describe electronic exchange and correlation. In order to assess the effect of different xc treatments additional calculations were performed with the local-density approximation (LDA) functional in the parameterization by Perdew and Wang \cite{Perdew1992} and with the hybrid functional B3LYP \cite{Becke1993,Stephens1994}. FHI-aims employs basis sets consisting of atom-centered numerical orbitals, and we specifically used the "tier2" basis set with the internal default tight settings for the numerical integrations. From test calculations at the "tier3" basis set level we conclude that calculated relative energies (and the ensuing $\Delta$SCF excitation energies) are converged to within 2\,meV at these numerical settings. The code also offers a standard implementation of the $\Delta$SCF approach \cite{Jones1989,Gunnarsson1976,Gorling1999} to obtain approximate excited states. The basis of this method are enforced modifications of the population of individual KS states, i.e. the KS equations are solved self consistently under the constraint of a given excited state occupation of the KS states, thereby accounting for orbital relaxation. To make full excited state PES scans and geometry optimizations tractable within $\Delta$SCF we modified this discrete constraint to a Gaussian smeared constraint that affects the population of all KS states within a defined small energy window of 0.01-0.02\,eV width. At generally insignificant changes of the total energy, this allows us to also readily converge systems with degenerate KS states that otherwise lead to significant problems in the SCF procedure. The actual geometry optimizations in both ground and excited states were then performed in combination with a locally modified version of the Atomic Simulation Environment (ASE) \cite{Bahn2002}. Forces were hereby relaxed to below a threshold value of 10 meV/{\AA}$^{-1}$.

In the most straightforward $\Delta$SCF realization, singlet excitations are mimicked by enforced population changes within one spin channel, while triplet excitations are modeled through appropriate occupation changes in both spin channels. In terms of the azobenzene frontier orbitals ({\em vide infra}), the singlet S1 state would thus result from the enforced occupation of the KS lowest unoccupied molecular orbital (LUMO) and the enforced depopulation of the KS highest occupied molecular orbital (HOMO) within one spin channel, while the triplet T1 state would result from the enforced occupation of the KS LUMO in one spin channel and the enforced depopulation of the KS HOMO within the other spin channel. However, earlier detailed work \cite{Gunnarsson1980,Ziegler1977,VonBarth1979,Behler2008} has demonstrated that the resulting single determinants of KS orbitals yield a particularly inaccurate description precisely of low-lying singlet states as are of central interest here. This owes to the fact that present-day local and semi-local xc functionals only evaluate the electron density that is symmetry-broken with respect to the true multi-determinantal singlet states. A possible remedy to this that we will employ throughout this work is the so-called "sum method" (SM) of Ziegler {\em et al.} \cite{Ziegler1977}, in which the multiplet corrected energy of the singlet state is calculated from the single-determinant singlet and triplet energies as $E^{\rm SM}_{\rm S} = 2 E_{\rm S} - E_{\rm T}$. A second possibility is to calculate a singlet state simply using non-spinpolarized DFT calculations, in which case the magnetization density is zero everywhere in space \cite{Behler2008}. In such restricted DFT calculations the occupations of the involved doubly-degenerate KS states are then simply varied by $\pm 1$. While this approach lacks a proper formal justification, it is particularly appealing in the context of the envisioned calculations for surface-mounted azobenzene, as there non-spinpolarized DFT would represent a significant saving in computational time.

The TD-DFT \cite{Bauernschmitt1996} and RI-CC2 calculations \cite{Christiansen1995,Hattig2000} used to assess the performance of the $\Delta$SCF approach were done with TURBOMOLE V6.2 \cite{Ahlrichs1989a}. TD-DFT calculations were hereby done for the same xc functionals as in the $\Delta$SCF case, as well as for the long-range corrected CAM-B3LYP functional \cite{Yanai2004}, which is presently not available in FHI-aims nor in TURBOMOLE. The CAM-B3LYP calculations were therefore performed with the GAMESS code\cite{Schmidt1993}. In addition to these excited state calculations TURBOMOLE V6.2 was employed to obtain the constrained PBE geometries for the PESs in Section \ref{PESchapter}. For these and the TD-DFT calculations we used a gaussian basis set of triple zeta quality (def2-TZVP) with polarization functions from the Ahlrichs series of basis functions \cite{Weigend1998}, and the resolution of identity (RI) approximation \cite{Bauernschmitt1997}. We estimate that the relative energies and excitation energies are converged within 10-20\,meV with respect to a quadruple zeta basis set. For the TD-DFT calculations the maximum value of the eucledian norm of the residual vector for the transition density matrices was set to 1$\cdot$10$^{-6}$. For the RI-CC2 calculations the basis set was def2-TZVPP, which was previously shown to yield highly accurate excitation energies for the azobenzene system \cite{Fliegl2003}. With respect to the higher def2-QZVPP basis set we estimate the uncertainty in the calculated relative and excitation energies to be about 60\,meV.

\section{Results and discussion}
\label{results}

\subsection{Isomerization of gas-phase azobenzene: Current state-of-the-art}

\begin{figure}
\includegraphics[width=\linewidth]{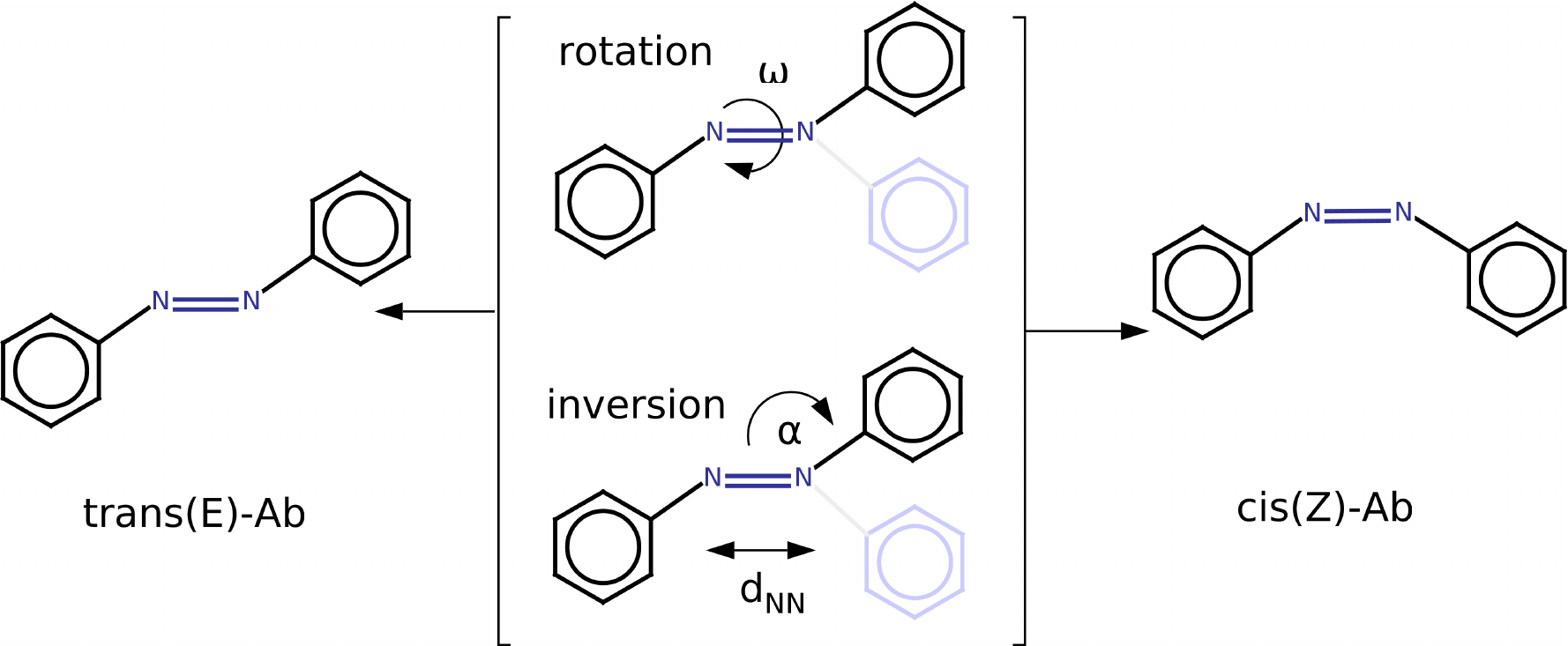}
\caption{\label{scheme}
Schematic overview of two possible photoisomerization mechanisms for azobenzene: rotation around the central CNNC dihedral angle $\omega$ (upper panel) and inversion around one of the NNC angles $\alpha$ (lower panel).}
\end{figure}
 
In order to unequivocally set the target for the approximate $\Delta$SCF approach we first briefly recapitulate the present state of understanding concerning the switching behavior of azobenzene and its derivatives, as it has emerged from a plethora of previous experimental and theoretical studies. This archetypical molecular switch is characterized by two stable isomers, namely \textit{cis} (Z) and \textit{trans} (E) azobenzene, whereby E-Ab is more stable by 0.6$\;$eV and both states are separated by a sizable ground-state barrier of about 1.6\,eV with respect to the E-Ab state \cite{Brown1975}. Both isomers are interconvertable via photoexcitation\cite{Rau2003}. Azobenzene can Z$\rightarrow$E or E$\rightarrow$Z isomerize following excitation in the UV-visible regime to either the so-called S1(n $\rightarrow\pi^*$) or the S2($\pi\rightarrow\pi^*$) state\cite{Rau2003}. The corresponding quantum yields for E$\rightarrow$Z (Z$\rightarrow$E) isomerization in \textit{n}-hexane are 0.40 or 0.12 (0.53 or 0.25) following excitation to S1 or S2, respectively\cite{Bortolus1979,Siampiringue1987}. Recently it has been shown that excitation to higher lying states predominantly leads to dissociation\cite{Bao2011}.

The detailed electronic and nuclear dynamics behind molecular Ab switching, especially photoinduced switching, have undergone a vast amount of experimental \cite{Rau1982,Bouwstra1983,Rau1984,Kumar1989,Lednev1996,Nagele1997,Fujino2002,Schultz2003,Satzger2003, Satzger2004,Chang2004,Bandara2010} and theoretical  \cite{Monti1982,Cattaneo1999,Ishikawa2001,Cembran2004,Fliegl2003,Diau2004, Gagliardi2004a,Tiago2005,Toniolo2005,Fuchsel2006,Crecca2006,Granucci2006,Conti2008,Yuan2008,Cusati2008,Shao2008,Ootani2009, Bockmann2008,Tiberio2010,Pederzoli2011,Weingart2011} investigation over the past decades and are still to some extent controversial \cite{Rau2003,Tamai2000,Fujino2002,Diau2004,Schultz2003}. Here, time resolved femtosecond absorption and fluorescence measurements pose particularly efficient tools to investigate the underlying electron dynamics from the experimental side. Braun and coworkers\cite{Satzger2004} measured transient absorption spectra and assigned several time constants to different parts of the isomerization process. For the S1 (S2) isomerization of E-Ab the authors find: $\tau_1$=0.34(0.42)$\;$ps, $\tau_2$=3.0(2.9)$\;$ps, $\tau_3$=12(12)$\;$ps. In the case of S2 excitation it was necessary to assign an additional smaller time constant of 0.13$\;$ps. From the similar time constants they concluded that both processes happen over fast nuclear motion from the Franck-Condon (FC) structure to the conical intersection (CI) ($\tau_1$), relaxation to the ground state minimum energy structure ($\tau_2$) and further vibrational cooling through the solvent ($\tau_3$). The authors interpret the additional process following S2 excitation as fast initial population transfer from S2 to S1. In this view, both excitation channels thus follow the S1 dynamics. For the isomerization starting from Z-Ab they reach the same conclusion even though the time constants show more subtle differences between S1 and S2 ($\tau_1$=0.17(0.2)$\;$ps, $\tau_2$=2(1.1)$\;$ps, $\tau_3$=10(14)$\;$ps). These experimental values have been confirmed by several other groups even though the corresponding authors came to different interpretations concerning the actual isomerization mechanism \cite{Nagele1997, Fujino2002, Schultz2003}. From their calculations of the involved PESs, Cattaneo and Persico\cite{Cattaneo1999} as well as Ishikawa and coworkers\cite{Ishikawa2001} also proposed that excitation to S2 is immediately followed by transitions that could involve several states and finally reach the S1 state, from whereon the actual isomerization then follows S1 dynamics. Recently this has been further supported by high level multireference calculations of Conti, Garavelli and Orlandi\cite{Conti2008} and explicit non-adiabatic dynamics simulations by Yuan and coworkers\cite{Yuan2008}.

For the nuclear S1 dynamics several possible pathways have been discussed in the literature. Figure \ref{scheme} illustrates the two most frequently studied mechanisms, namely an isomerization around the central CNNC dihedral angle (''rotational pathway'') and an isomerization around one of the two CNN angles (''inversion pathway''). The general understanding of the prevalence of these mechanisms has undergone various transitions. The initial belief was that excitation to S1 mainly follows inversion whereas excitation to S2 should follow rotational isomerization\cite{Monti1982,Rau2003}, this way rationalizing the different quantum yields. However, most of the recent experimental studies \cite{Satzger2003, Satzger2004, Schultz2003}, as well as theoretical studies that either investigated the excited PESs \cite{Ishikawa2001,Cembran2004,Tiago2005,Conti2008} or performed explicit non-adiabatic dynamics simulations \cite{Toniolo2005,Granucci2006,Yuan2008,Shao2008,Ootani2009,Bockmann2010} agree on the dominance of the rotational isomerization following excitation in either S1 or S2 for azobenzene in gas-phase and solvent.

\subsection{Ground and excited state PES topology}
\label{PESchapter}

 \begin{figure*}
 \includegraphics[width=17cm]{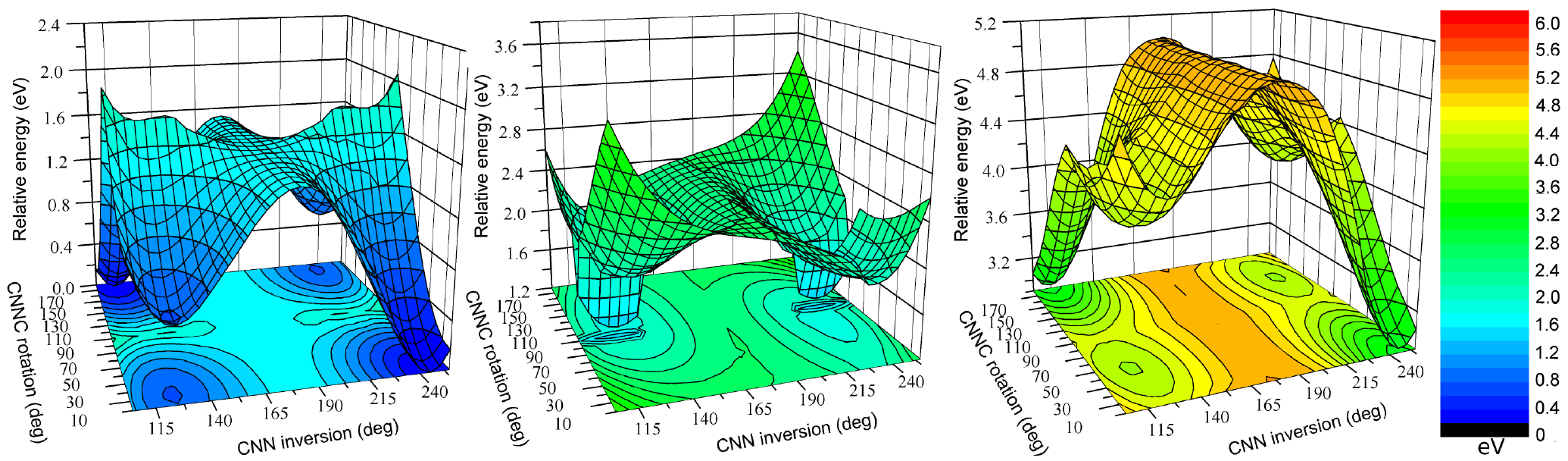}
 \caption{\label{DSCF-PES} Two-dimensional relaxed $\Delta$SCF(GGA-PBE) PES scans of rotation around the dihedral CNNC angle $\omega$ and inversion around one of the two CNN angles $\alpha$, cf. Fig. \ref{scheme}. Shown are the ground state S0 (left), the first excited state S1 (center) and the second excited state S2 (right). Energies relative to the zero reference E-Ab ground state energy are given in eV.}
 \end{figure*}

As established in preceding detailed quantum chemical work the centrally targeted low-lying excitations, S1 and S2, have largely singly excited character at least at the FC structures, and can be viewed as $n \rightarrow \pi^*$ (HOMO to LUMO) and $\pi\rightarrow\pi^*$ (HOMO-1 to LUMO) transitions, respectively \cite{Fliegl2003,Conti2008}. In SM-corrected $\Delta$SCF we accordingly model the two singlet excitations by modifying the populations of HOMO and LUMO (S1) and HOMO-1 and LUMO (S2), respectively. Figure \ref{DSCF-PES} shows correspondingly obtained two-dimensional PES scans along the dihedral CNNC angle $\omega$ and along one of the two CNN angles $\alpha$, cf. Fig. \ref{scheme}. Each point of the PES corresponds to a molecular geometry, in which the values for these two angles were constrained to the specific value, while all other degrees of freedom of the azobenzene molecule are those as resulting from a full geometry optimization in the ground-state. Rather than a state-specific geometry optimization, this allows to clearly disentangle geometric and electronic effects and thereby to directly compare different methods as all are evaluated for the same geometry (also in Fig. \ref{1DPES} below). The PESs shown in Fig. \ref{DSCF-PES} are for the GGA-PBE xc functional, and we obtain essentially the same topologies for the three surfaces with the LDA or B3LYP. The only difference are more or less constant offsets between the three surfaces depending on the level of xc treatment, which is why we restrict the presentation for the moment to the GGA-PBE case and return to the xc discussion in the next section when focusing on the vertical excitation energies.

Qualitatively, the overall obtained topology of S0, S1 and S2 is perfectly consistent with the prevalent understanding of the azobenzene photochemistry as summarized in the preceding section. The ground-state PES is dominated by the two metastable states, E-Ab and Z-Ab, separated by sizable barriers along both the rotation and inversion pathway. In contrast, the S1 PES does not exhibit a barrier along the rotational pathway, which after photoexcitation of either E-Ab or Z-Ab should thus quickly lead the system to the well-known CI region around mid-rotation. Moreover, the S1-FC region at E-Ab is rather flat, while the S1-FC region at Z-Ab is very steep. This is perfectly consistent with the experimentally reported longer excited state lifetime of E-Ab compared to Z-Ab, and is also in line with the reported lower S1 quantum yields for E$\rightarrow$Z than for Z$\rightarrow$E isomerization. Finally, the closeness of the S2 minima to the respective S0-S2 FC structures, as well as the separation of these minima by large barriers along both inversion and rotation suggests that isomerization after S2 excitation does indeed not occur on the S2 surface, but rather via deexcitation along CIs in other degrees of freedom than those scanned here, followed by motion on the S1 surface.

 \begin{figure*}
 \includegraphics[width=\linewidth]{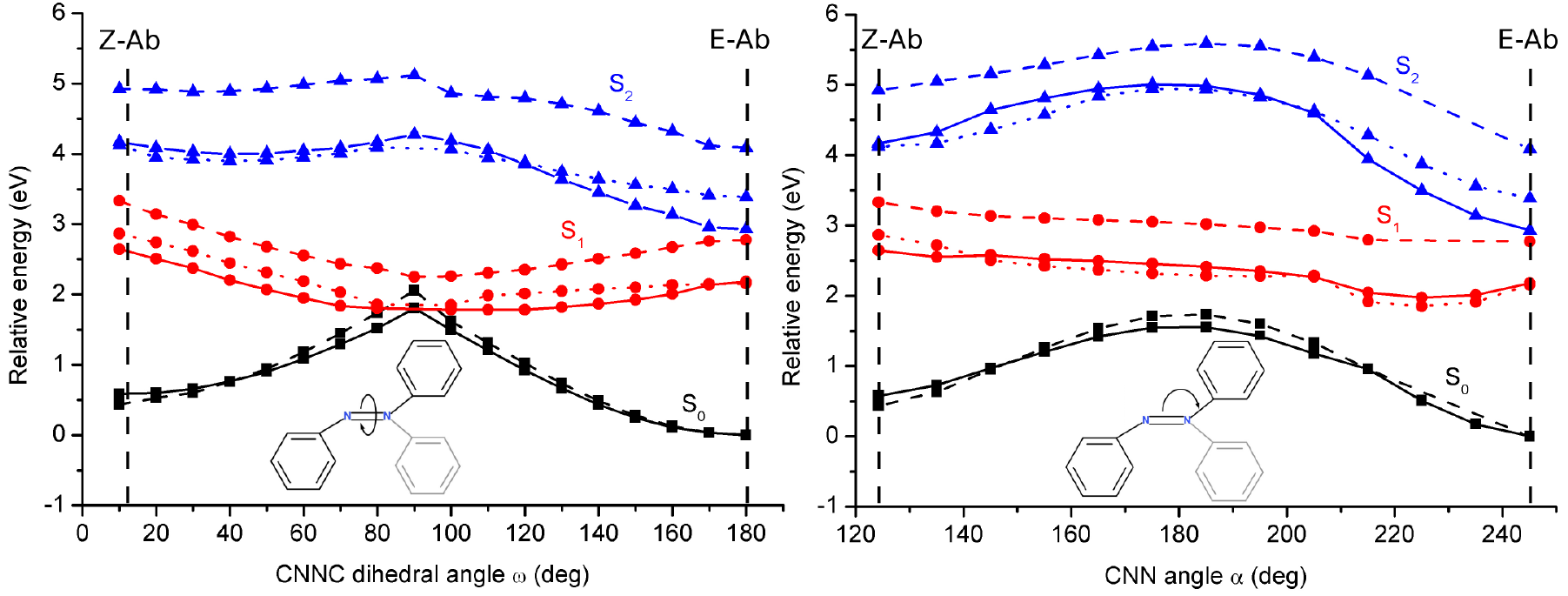}
 \caption{\label{1DPES} PES scans along the rotational (left) and inversion (right) pathway, cf.  Fig. \ref{scheme}. Shown are the groundstate (S0, black), first (S1, red) and second (S2, blue) excited states, calculated each time with $\Delta$SCF(GGA-PBE) (solid line), TDDFT(GGA-PBE) (dotted line) and RI-CC2 (dashed line).}
 \end{figure*}
 
Essentially the same S0, S1 and S2 topologies are also obtained at the RI-CC2 and TD-DFT level of theory, with the TD-DFT PESs obtained at different xc functional levels again primarily vertically shifted against each other. We present the corresponding two-dimensional PESs in the supplementary material\cite{supplement} and focus here instead on one-dimensional PES scans which allow for a more quantitative comparison. Figure \ref{1DPES} compiles these scans along the two prevalently discussed isomerization pathways, the rotational one following motion around the dihedral angle $\omega$ and the inversion one following motion along one of the two CNN angles $\alpha$, cf. Fig. \ref{scheme}. The basis of these one-dimensional PES scans are again optimized ground-state geometries, in which the corresponding angle was frozen and all other degrees of freedom were fully relaxed.

The topological similarity, i.e. relative energetics within each PES, for the three methods is rather striking. At the ground-state it is reflected by an almost quantitative agreement of the inversion barrier computed with DFT(GGA-PBE) and RI-CC2, 1.52\,eV and 1.74\,eV, respectively. These values match also nicely with experiment and previously reported values of 1.5\,eV (DFT(LDA) \cite{Tiago2005}) and 1.74\,eV (DFT(B3LYP) \cite{Crecca2006}). Only slightly lower quantitative consistency is achieved for the S0 rotation barrier, for which we compute values of 1.81\,eV and 2.05\,eV at DFT(GGA-PBE) and RI-CC2 level of theory, while values of 1.65\,eV (CASSCF/PT2 \cite{Cembran2004}) and 2.18\,eV (DFT(B3LYP) \cite{Crecca2006}) can be found in literature. At the RI-CC2 reference, the rotation pathway is dominated by the well known CI seam between S0 and S1 state around midway rotation \cite{Cembran2004, Diau2004, Tiago2005, Crecca2006}. The residual gap obtained in the present calculations, cf. Fig. \ref{1DPES}, is hereby due to the fact that the geometries along the scan were optimized at the DFT level. For the inversion pathway RI-CC2 reveals no intersection between S0 and S1 along the here displayed PES cut. Also this is in agreement with preceding work, which either did not find any near-degeneracies between states on the inversion pathway or found them only at very high energies compared to the CI seam on the rotation pathway \cite{Cembran2004}. In both cases, i.e. rotation and inversion pathway, $\Delta$SCF(GGA-PBE) and TD-DFT(GGA-PBE) correctly reproduce the existence viz. non-existence of S1 and S0 state degeneracies, cf. Fig. \ref{1DPES}. At the S2 surface, the large barrier along the inversion pathway is again rather well reproduced by the three methodologies, 2.08\,eV ($\Delta$SCF(GGA-PBE)), 1.55\,eV (TD-DFT(GGA-PBE)), 1.50\,eV (RI-CC2) when measured from E-Ab, or 0.84\,eV ($\Delta$SCF(GGA-PBE)), 0.82\,eV (TD-DFT(GGA-PBE)), 0.67\,eV (RI-CC2) when measured from Z-Ab. The same holds for the barrier along the rotation pathway on S2, where the corresponding values are 1.35\,eV ($\Delta$SCF(GGA-PBE)), 0.71\,eV (TD-DFT(GGA-PBE)), 1.03\,eV (RI-CC2) when measured from E-Ab.

 \begin{table*}
 \caption{\label{optimization} Optimized geometry parameters of E and Z azobenzene in ground (S0) and excited (S1,S2) states, cf. Fig. \ref{scheme} for the definition of the azo-bridge bond length $d_{\rm NN}$ and the two angles $\omega$ and $\alpha$. Additionally shown are the relative energies $\Delta E$ of the corresponding states with respect to the ground-state E-Ab zero reference. None of the methods identified a stable minimum after optimization from S1 Z-Ab, which is why the corresponding entries have been left blank in the table.}
 \begin{tabular}{ccccccccccc} \hline
   &       method  &  \multicolumn{4}{c}{Trans (E)}     & & \multicolumn{4}{c}{Cis (Z)}   \\  \cline{3-6} \cline{8-11}
   &             &  $\omega$  &  $\alpha$   & d$_{\rm NN}$ & $\Delta E$ & &  $\omega$  &  $\alpha$  & d$_{\rm NN}$ & $\Delta E$ \\ 
   &             &  \multicolumn{2}{c}{(deg)} & (\AA) &   (eV)    & &  \multicolumn{2}{c}{(deg)} & (\AA) &  (eV) \\ \hline
S0 & DFT(GGA-PBE)              & 180  & 115     & 1.26 & 0    & \hspace{0.4cm} & 12   &  124 & 1.25 & 0.58    \\ 
   & RI-CC2                    & 180  & 114     & 1.27 & 0    & & 7    &  121 & 1.27 & 0.47    \\
   & CASSCF/PT2\cite{Cembran2004}     &180 & 115& 1.24 & 0    & & 4    &  123 & 1.24 & 0.52    \\
   & Exp\cite{Bouwstra1983,Mostad1971}&180 & 114& 1.25 & 0    & & 0    &  122 & 1.25 & 0.6\cite{Brown1975}   \\
S1 & $\Delta$SCF(GGA-PBE)      & 180  & 130     & 1.25 & 1.67 & & $-$  &  $-$ & $-$ & $-$   \\ 
   & TD-DFT(GGA-PBE)           & 180  & 131     & 1.24 & 1.53 & & $-$  &  $-$ & $-$ & $-$     \\
   & RI-CC2                    & 180  & 128     & 1.26 & 2.26 & & $-$  &  $-$ & $-$ & $-$     \\
   & CASSCF/PT2\cite{Cembran2004}  & 180  & 129 & 1.25 & 1.95 & & $-$  &  $-$ & $-$ & $-$     \\
S2 & $\Delta$SCF(GGA-PBE)      & 180  & 113     & 1.36 & 2.53 & & 18   &  127 & 1.31 & 3.60   \\ 
   & TD-DFT(GGA-PBE)           & 180  & 111     & 1.34 & 3.15 & & 30   &  122 & 1.31 & 3.65   \\
   & RI-CC2                    & 180  & 110 /113& 1.37 & 4.06 & & $-$  &  $-$ & $-$ & $-$     \\ 
   & CASSCF/PT2\cite{Conti2008}& 180  & 113     & 1.35 & 4.05 & & 8    &  129 & 1.29 & 5.55   \\
\hline
 \end{tabular}
 \end{table*}

We find the excellent agreement of the three methods ($\Delta$SCF, TD-DFT, RI-CC2) with respect to the topology to also extend to other parts of the PES not contained in the hitherto presented scans. This is nicely demonstrated by Table \ref{optimization}, which compiles selected structural parameters of the ground-state E-Ab and Z-Ab states, as well as of those minimum energy structures on the S1 and S2 states that are obtained after optimization from the E-Ab and Z-Ab FC structure. At the ground-state level we reproduce the known excellent performance of the DFT GGA-PBE functional in describing both geometry and relative energetics of both E-Ab and Z-Ab isomers as compared to both higher-level theory and experiment. Optimization in the S1 state with FC E-Ab as starting structure yields a minimum at very similar geometries within $\Delta$SCF(GGA-PBE), TD-DFT(GGA-PBE) and RI-CC2, which in turn compare nicely to the CASSCF optimized geometry reported by Cembran {\em et al.} \cite{Cembran2004}. The $d_{\rm NN}$ bond length at this minimum energy geometry is essentially unchanged with respect to ground-state E-Ab, as one would intuitively expect for a transition depopulating the non-bonding $n$ HOMO orbital. In contrast, in terms of energetics the different methodologies yield again a large scatter and we will return to this point in the next section. In the S2 state an equivalent situation is obtained. Excellent agreement in the geometries is contrasted by strong discrepancies in the energetics for both minimum energy structures obtained after optimization from the FC E-Ab and FC Z-Ab. With respect to the angular degrees of freedom both these minimum energy structures exhibit values in close correspondance to their S0 counterparts. They differ largely in their strongly activated $d_{\rm NN}$ bond though. This is again congruent with the expectations for a transition depopulating the $\pi$-bonding HOMO-1 orbital and supports the already mentioned interpretation that deexcitation from S2 occurs via CIs in other degrees of freedom than those relevant for the isomerization process.

\subsection{Vertical excitation energies}

 \begin{figure*}
 \includegraphics[width=\linewidth]{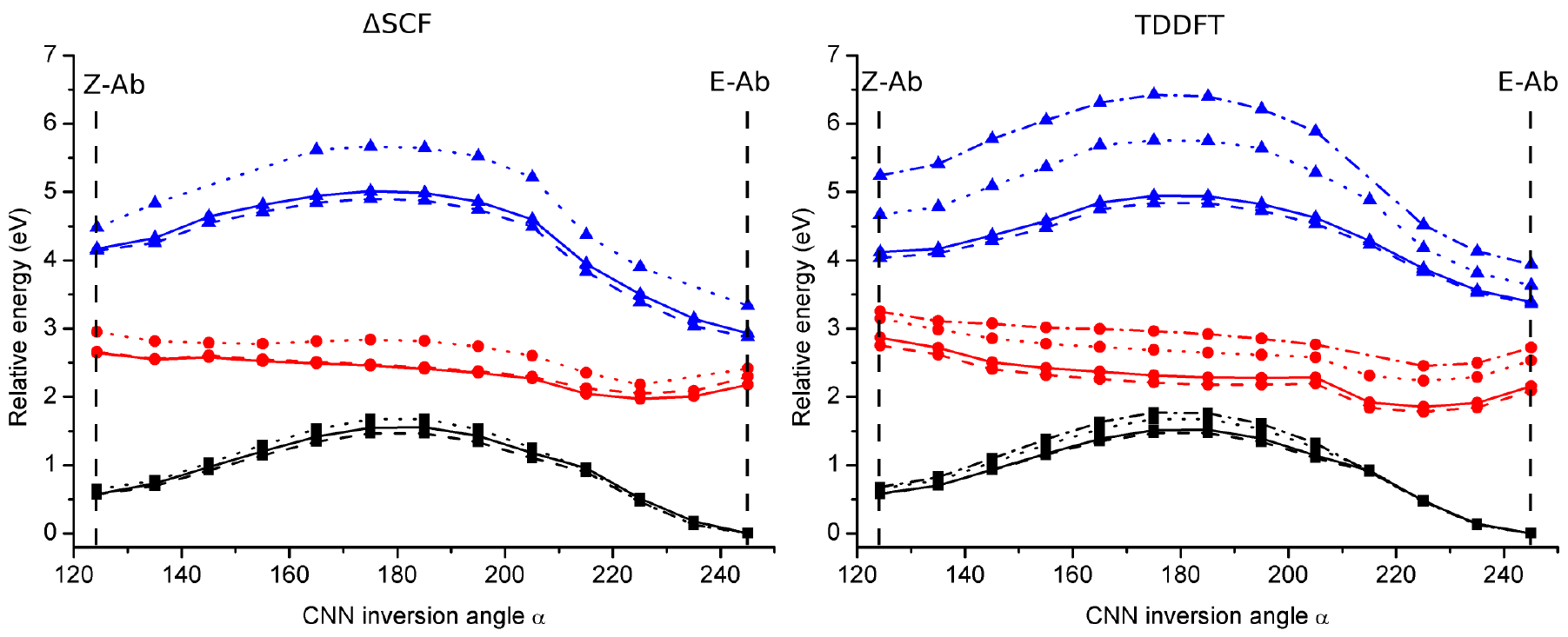}
 \caption{\label{1DPES2} PES scan along the inversion pathway, calculated with $\Delta$SCF (left side) and TDDFT (right side). Shown are the groundstate (S0, black), first (S1, red) and second (S2, blue) excited states, calculated each time with an LDA (dashed line), GGA-PBE (straight line) and B3LYP (dotted line) functional. Also shown for the TDDFT case is the pathway calculated with the CAM-B3LYP functional (dashed-dotted line).}
 \end{figure*}
 
\begin{table}
\caption{\label{vertical} Vertical excitation energies for S1 and S2 excitation at E-Ab and Z-Ab at the different levels of theory and from experiment.}
\begin{center}
\begin{tabular}{cccccc}
\hline method & \multicolumn{2}{c}{Trans (E)} & \hspace{0.4cm} &\multicolumn{2}{c}{Cis (Z)}  \\ \cline{2-3} \cline{5-6}
  S0$\rightarrow$ & S1 & S2 & & S1 & S2 \\
& \multicolumn{2}{c}{(eV)} & & \multicolumn{2}{c}{(eV)} \\ \hline
$\Delta$SCF(LDA) & 2.30 & 2.88 & & 2.10 & 3.56 \\
TD-DFT(LDA) & 2.09 & 3.36 & & 2.19 & 3.46 \\ \hline
$\Delta$SCF(PBE)-not spin pol.& 2.27 & 2.75 & & 2.13 & 3.54  \\
$\Delta$SCF(PBE) & 2.21 & 2.98 & & 2.10 & 3.63 \\
TD-DFT(PBE) & 2.15 & 3.39 & & 2.29 & 3.43 \\ \hline
$\Delta$SCF(B3LYP)            & 2.41 & 3.33 & & 2.30 & 3.83 \\
TD-DFT(B3LYP) & 2.53 & 3.63 & & 2.49 & 4.01 \\ \hline
TD-DFT(CAM-B3LYP) & 2.72 & 3.94 & & 2.58 & 4.56 \\ \hline
RI-CC2 & 2.84 & 4.07 & & 3.00 & 4.51 \\
CASSCF/PT2 \cite{Conti2008} & 2.53 & 4.23 &  & 2.72 & 4.49 \\ \hline
Exp. \cite{Andersson1982,Biancalana1999}     & 2.82 & 4.12 &  & 2.92 & 4.68 \\ \hline
\end{tabular}
\end{center}
\end{table}

As already mentioned in the preceding section, equivalent S0, S1 and S2 topologies to those just discussed are obtained when using different xc functionals in the DFT, $\Delta$SCF and TD-DFT calculations. {\em Cum grano salis} corresponding PESs essentially exhibit global vertical shifts with respect to each other as further illustrated in Fig. \ref{1DPES2} for the inversion pathway. This allows us to focus the discussion on the different levels of theory on single prominent points on the PES, suitably the vertical excitation energies as there experimental data is also available as reference. Table \ref{vertical} summarizes the corresponding data. The benchmark against the experimental vertical excitation energies at E-Ab and Z-Ab emphasizes the known high accuracy achieved by the RI-CC2 approach for the two low-lying Ab singlet excitations \cite{Fliegl2003} and justifies its use as a theoretical reference method in our study. In contrast, TD-DFT and $\Delta$SCF based on present-day local and semi-local functionals yield excitation energies that are dramatically too low, as had already been noticed in preceding work for this molecule \cite{Fliegl2003,Tiago2005}. This concerns predominantly the S2 excitation which at Z-Ab is underestimated by more than 1\,eV.

While thus unsatisfactory on the absolute scale, the similarity of the results produced by $\Delta$SCF and TD-DFT as long as they are based on the same functional is notable. The differences are with $\sim 0.4$\,eV largest for S2 E-Ab, while for the other three excitations listed in Table \ref{vertical} the two methods match to within 0.1-0.2\,eV. To one end this is due to the sum-rule correction we employ for $\Delta$SCF, which yields a spin purified state that is better comparable to TD-DFT \cite{Kowalczyk2011}. The plain spin-mixed $\Delta$SCF approach instead yields excitation energies for the two states that are typically about 0.3\,eV lower than the spin-purified ones. This would further increase the difference to the corresponding TD-DFT values and, worse, also to experiment. Table \ref{vertical} furthermore indicates that simple non-spin polarized $\Delta$SCF calculations \cite{Behler2008} are also in this system an alternative, effective way of tackling the multi-determinantal singlet problem. The corresponding values do not differ much from the sum-method corrected ones, and come at a significantly lower computational cost.

To the other end the obtained similarity of $\Delta$SCF and TD-DFT results is connected to the pronounced single-particle character of the S1 and S2 excitations, which in turn is also the rationalization for the high accuracy of the RI-CC2 method \cite{Fliegl2003}. This is most obvious for the S1 state, which is essentially described by a single excitation over the entire PES range scanned and which is thus most straightforwardly mimicked in $\Delta$SCF. However, it also holds to some extent for the S2 state, which is of a more collective nature in the sense that it exhibits multiple significant TD-DFT excitation amplitudes. As apparent from the afore described not excessively large deviations even this can still be relatively well described by single-excitation $\Delta$SCF due to its desirable ability to account for orbital relaxation ({\em vide infra}).

Despite the similarity, the absolute performance of both methods when based on local and semi-local functionals against RI-CC2 and experiment is still a concern. Particularly the deviation in the S2 Z-Ab vertical excitation energy exceeds the one commonly found for low-lying singlet excitations in organic molecules \cite{Kowalczyk2011,Peach2008,Jacquemin2009}. Visual inspection of the involved frontier orbitals, cf. Fig. \ref{Orbitale} below, suggests that this might be related to some charge-transfer (CT) aspect of the excitations, which in particular for S2 around Z-Ab shifts charge between the central azo-group and the phenyl-moieties. This interpretation receives some quantitative support by an evaluation of Tozer's CT $\Lambda$-parameter \cite{Peach2008}. Measuring the spatial overlap in a given excitation, $\Lambda$ values towards unity indicate that occupied and virtual orbitals involved in
the excitation occupy increasingly similar regions of space. In contrast to Rydberg and CT excitations such "short-range excitations" should then be much better amenable to TD-DFT based on local or semi-local xc functionals \cite{Peach2008,Silva-Junior2008,Casida2009}.
For the present case, it is indeed precisely for S2 and around Z-Ab that we obtain the lowest $\Lambda$-values around 0.5, while for S1 and for S2 towards E-Ab $\Lambda$-values lie consistently around 0.7 or higher. Further support for the CT picture comes then also from the larger reduction of the underestimation of S2 at Z-Ab when going to the hybrid functional level. Whereas for S1 and S2 at E-Ab the admixture of exact exchange reduces the error with respect to the RI-CC2 from $\sim 0.7$\,eV to $\sim 0.4$\,eV, at Z-Ab the S2 error goes from the larger $\sim 1.1$\,eV equally down to $\sim 0.5$\,eV, cf. Table \ref{vertical}. At the level of the Coulomb-attenuated CAM-B3LYP functional this remaining deviation is then further reduced to the order of 0.1\,eV throughout. Also this is consistent with the interpretation of some overall CT character of the low-lying singlet excitations, which this functional with its varying degree of exact exchange at short and long range is specifically supposed to tackle \cite{Tozer1998, Casida1998, Allen2000, Gruening2001}. At this functional level the agreement reached with respect to experiment is thus essentially {\em en par} with that of the correlated wave function reference techniques, and we speculate that the same would approximately hold for CAM-B3LYP-based $\Delta$SCF (which unfortunately is presently not available to us).

\subsection{TD-DFT and inversion path barrier}

 \begin{figure*}
 \includegraphics[width=0.7\linewidth]{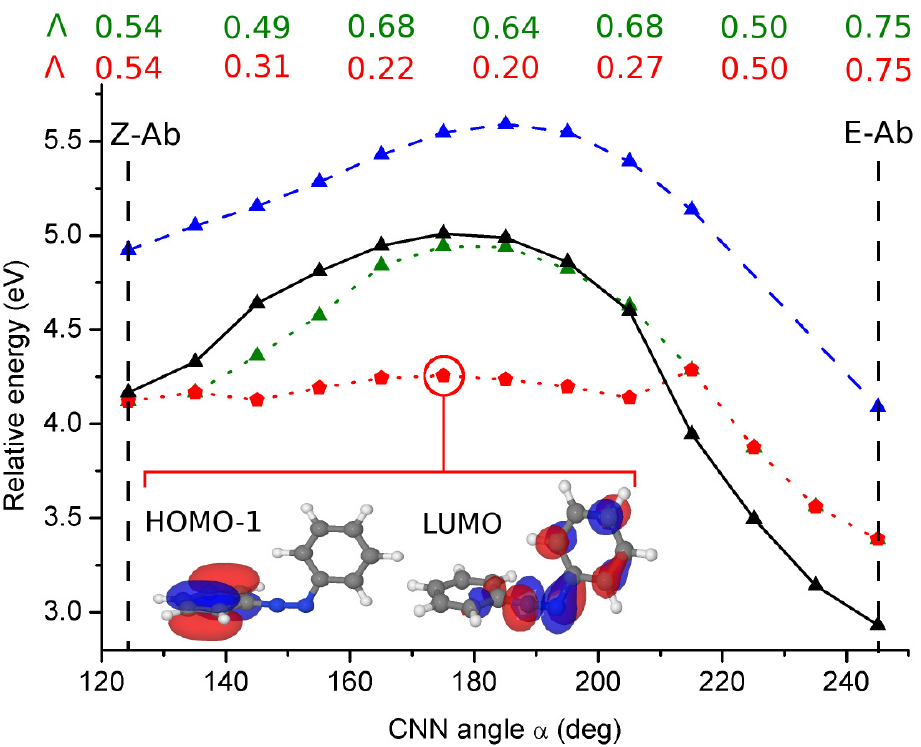}
 \caption{\label{Lambda} Scan along the inversion pathway, comparing the S2 PES as calculated with TD-DFT(GGA-PBE) when following the HOMO-1$\rightarrow$LUMO excitation (red, dotted line) and when following the transition with predominant $\pi\rightarrow\pi^*$ character (green dotted line). Only the latter approach yields the correct PES topology with sizable barrier around mid-inversion as compared to RI-CC2 (dashed line) and $\Delta$SCF(GGA-PBE) (solid line). Also shown are the CT $\Lambda$-values \cite{Peach2008} for the two TD-DFT curves (see text), as well as the LUMO and the "wrong" HOMO-1 orbital at mid-inversion. The latter nicely reveal the pronounced CT-character of this transition.}
  \end{figure*}

The good agreement of $\Delta$SCF, TD-DFT and RI-CC2 in terms of overall PES topology particularly around mid-inversion might come as a bit of a surprise in view of earlier studies that reported significant discrepancies for this \cite{Tiago2005} or for $\pi$-bond twisting paths of comparable molecules \cite{Wanko2004,Wiggins2009,Ploetner2010}. In such cases deviations between $\Delta$SCF and TD-DFT are often readily attributed to the "simplicity" of the prior theory. Alternatively, "collective" character of an excitation as judged from the existence of several significant TD-DFT amplitudes is also cited as reason for the failure of "single excitation" restricted $\Delta$SCF. In turn, when it comes to differences between TD-DFT and higher level wave function theories, the deficiency of semi-local TD-DFT to deal with CT-character of excitations is a frequently encountered rationalization. Instead, in the present case yet another difficulty of TD-DFT applies, namely state-crossings, and we find it instructive to point out that this can easily lead to wrong assessments, in particular as we find the allegedly "simpler" theory $\Delta$SCF to be significantly more robust with respect to this issue.

 \begin{figure*}
 \includegraphics[width=\linewidth]{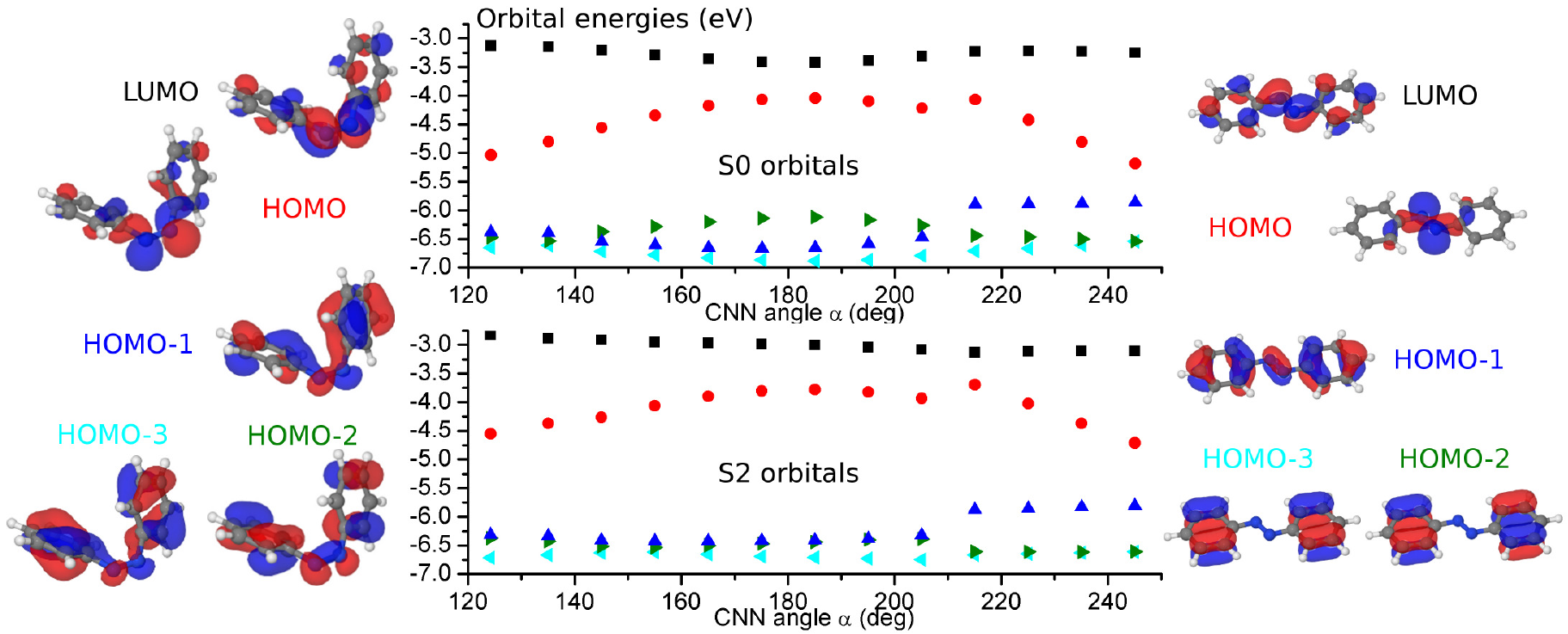}
 \caption{\label{Orbitale} Energetic positions of GGA-PBE KS frontier orbitals along the inversion pathway as resulting from a self-consistent ground-state calculation (upper panel) and as resulting from a self-consistent $\Delta$SCF calculation for the S2 excitation (lower panel). Additionally shown are the corresponding KS ground-state orbital shapes at Z-Ab (left) and E-Ab (right).}
 \end{figure*}

The reason why a similar S2 topology with sizable barrier around mid-inversion was found for $\Delta$SCF, TD-DFT and RI-CC2 in Fig. \ref{1DPES} is that we do not simply plot the values following the transition from the second highest occupied to the lowest unoccupied orbital without considering the orbital character in the case of TD-DFT. Instead, we always specifically follow the excitation that exhibits the largest amplitude for the targeted transition between the $\pi$ and $\pi^*$ KS orbitals. As shown in Fig. \ref{Lambda} the two procedures do not lead to the same result around mid-inversion, and only the approach that tracks the correct transition yields a PES topology that is in agreement with both $\Delta$SCF and the reference RI-CC2 data. In contrast, the approach that merely monitors the HOMO-1 to LUMO TD-DFT excitation yields a very wide plateau-region along the inversion path, precisely as reported previously by Tiago {\em et al.} \cite{Tiago2005} for this system. 

The source for this difference is clearly apparent from Fig. \ref{Orbitale}, which shows the evolution of the energetic position of the GGA-PBE KS frontier orbitals along the inversion pathway. At both E-Ab and Z-Ab these orbitals exhibit the ordering HOMO-1 ($\pi$), HOMO ($n$) and LUMO ($\pi^*$) as intuitively expected from the $n \rightarrow \pi^*$ (HOMO to LUMO) and $\pi\rightarrow\pi^*$ (HOMO-1 to LUMO) character of S1 and S2, respectively. However, around mid-inversion the ground-state KS orbital energies of the states that at E-Ab and Z-Ab correspond to HOMO-1 and HOMO-2 cross. Inspection of the TD-DFT amplitudes reveals that the correct $\pi\rightarrow\pi^*$ TD-DFT excitation then corresponds to a transition primarily between this HOMO-2 level and the LUMO. 

Here, it is intriguing to note that also in Hartree-Fock as the basis for the RI-CC2 reference, the frontier orbital ordering does not correspond to the aforementioned $\pi$, $n$, and $\pi^*$ sequence. Still, the two lowest energy RI-CC2 excitations anywhere on the PES parts scanned in this study have predominant amplitude just exactly for the transitions expected, i.e. $n \rightarrow \pi^*$ for S1 and $\pi\rightarrow\pi^*$ for S2. As such we interpret the TD-DFT result which places the $\pi\rightarrow\pi^*$ transition at a higher lying excitation than the second transition around mid-inversion as the inability of linear-response theory to cope with the bad ground-state orbitals offered by GGA-PBE. This is then also consistent with the observation that in TD-DFT the S2 excitation typically exhibits larger amplitudes for more than one single-particle transition \cite{supplement}.

Intriguingly, the allegedly simpler theory $\Delta$SCF is much less affected by this limitation as it allows for orbital relaxation under the excitation constraint. Figure \ref{Orbitale} demonstrates that the self-consistent orbitals obtained under the S2 population constraint no longer exhibit any state crossing along the inversion pathway. Once self-consistency is achieved, the constraint of a depopulated HOMO-1 and a populated LUMO leads always to a situation where the HOMO-1 corresponds to the $\pi$ state and the LUMO to the $\pi^*$ state, i.e. the computed excitation energy corresponds indeed exactly to the transition we wanted to model. We found this type of robustness to hold for both S1 and S2, everywhere on the PES parts we scanned, and for whatever xc functional we used. Especially for dynamical simulations or mappings of larger parts of the PESs this is in our eyes an important asset. It is also particularly remarkable as at the hybrid functional level (B3LYP and CAM-B3LYP) the ordering of the ground-state KS levels differs from expected $\pi$, $n$, $\pi^*$ sequence essentially everywhere on the PES parts we scanned. This made it rather cumbersome to track the correct transitions in TD-DFT, which, however, was the prerequisite to obtain the consistent agreement of the TD-DFT PES topologies with respect to $\Delta$SCF and RI-CC2 reported above. For corresponding method comparisons the ease with which unidentified state-crossings can impair the TD-DFT results is hereby particularly consequential, as it may readily lead to wrong assessments. Discrepancies between TD-DFT and $\Delta$SCF that in reality are due to an unidentified state-crossing in TD-DFT may lead to the dismissal of the allegedly "simpler" $\Delta$SCF theory \cite{Tiago2005}. As shown in Fig. \ref{Lambda} the switch of the excitation character induced by the state-crossing around mid-inversion gives furthermore rise to small $\Lambda$ values for the wrong TD-DFT transition. This bears the danger to assign the discrepancy in the PES topology of this transition with respect to the RI-CC2 reference incorrectly to the deficiency of present-day functionals in describing CT excitations.

\section{Conclusion and outlook to surface-mounted azobenzene}
\label{conclusion}

In summary we have systematically computed ground- and low-lying singlet excited state PESs that are of relevance for the isomerization dynamics of gas-phase azobenzene. Our results demonstrate that sum-method corrected $\Delta$SCF yields global PES topologies, i.e. relative energetics within one PES, that agree very well with those of TD-DFT at the same xc functional level and with accurate RI-CC2 reference data. Previous contradictory reports concerning the agreement of $\Delta$SCF and TD-DFT suffered from unresolved state crossings in the TD-DFT calculations \cite{Tiago2005}, while the orbital relaxation possible in $\Delta$SCF makes this approach very robust with respect to this issue. The now unanimously obtained PES topologies of S0, S1 ($n \rightarrow \pi^*$) and S2 ($\pi \rightarrow \pi^*$) states are furthermore quite consistent with existing experimental data concerning the photo-isomerization mechanism.

When based on the same xc functional sum-method corrected $\Delta$SCF and TD-DFT agree to within 0.1-0.2\,eV for the S1 state that is most relevant for the isomerization of free azobenzene, and to within 0.4\,eV for the S2 state. This suggests $\Delta$SCF as a promising route to larger azobenzene-containing systems, where TD-DFT becomes computationally untractable. This concerns predominantly surface-mounted azobenzene, where preliminary calculations for the adsorption at coinage metals indicate that the low lying excited states largely retain their molecular character \cite{McNellis2009}. Particularly appealing in this context is that we obtain the correct PES topology for gas-phase azobenzene already at the level of semi-local xc functionals, which are presently the unbeaten workhorse for metal adsorption studies.

Notwithstanding, at this level of xc treatment the vertical excitation energies produced by $\Delta$SCF and TD-DFT are grossly underestimated. Our analysis attributes this primarily to some charge-transfer character of the S1 and S2 excitations, which the local functionals are unable to grasp. As to be expected hybrid (B3LYP) and even more so long-range corrected hybrid (CAM-B3LYP) functionals improve on this situation. Conserving the overall topology they primarily induce global upward shifts of the excited state PESs, leading to vertical excitation energies at the CAM-B3LYP level that are roughly {\em en par} to the correlated wave function reference techniques. 

While the reduction of self-interaction error achieved with the advanced functionals is therefore of paramount importance to adequately describe the molecular excitations, one has to recognize that mere admixture of exact exchange does not seem to be the right pathway for adsorption at metal surfaces (at least when judged from the few seminal ground-state studies performed to date.\cite{Hu2007,Carter2008,Stroppa2008}). There, semi-local functionals are still more or less the only tractable approach. For future applications to surface-mounted azobenzene this dictates a cautious approach carefully assessing what can and what cannot be addressed with semi-local functional based $\Delta$SCF. The present results give good hope that the qualitative PES topology may be retrieved, with recent methodological $\Delta$SCF extensions \cite{Hellman2004,Gavnholt2008} of particular interest in case of appreciable adsorbate-surface hybridization. In the best of cases this is already sufficient to conclude on the isomerization mechanism. As to the absolute excitation energies the situation is less clear. Semi-local levels of xc treatment that are adequate for the extended metallic states will be insufficient for the localized molecular states. We speculate that an insightful mixture of extended supercell surface calculations and suitable molecular model system reference calculations might nevertheless yield highly valuable insight to this end.

\begin{acknowledgments}
Funding through the German Research Council, as well as the support of TUM Graduate School's Faculty Graduate Center Chemistry is gratefully acknowledged. We thank Joerg Meyer and Christoph Scheurer for fruitful discussions and skillfull technical support, as well as Eva Deront for supplementary calculations. CPU time was generously provided at HLRB-II of the Leibniz Rechenzentrum.
\end{acknowledgments}

\end{document}